\documentclass[twocolumn,mnras,aps,superscriptaddress,preprintnumbers,showpacs,nofootinbib,eqsecnum,amsfonts,amsmath,linenumbers]{aa}  

\usepackage{graphicx}
\usepackage{txfonts}
\usepackage[switch]{lineno}
\usepackage{amsmath}
\usepackage{amssymb}
\usepackage{graphicx}
\usepackage{xspace}
\usepackage{xcolor}
\usepackage{orcidlink}
\newcommand{\gwo}{\texttt{GWEMOPT}\xspace}
\newcommand{\lcm}{\texttt{lightcurve-matching}\xspace}

\begin{document}

   \title{Optical Follow-Up Strategies for the Next Neutrino-Detected Galactic Core-Collapse Supernova}


   \author{
        P.-A.~Duverne \inst{1} \orcidlink{0000-0002-3906-0997}
        \and
        W.~K.~Mouici \inst{1}
        \and
        A. Coleiro \inst{1} \orcidlink{0000-0003-0860-440X}
        \and 
        J.-G.~Ducoin \inst{2} \orcidlink{0009-0008-7341-4825}
        \and
        M.~W.~Coughlin \inst{3} \orcidlink{0000-0002-8262-2924}
        }

    \institute{
            Universit\'e Paris Cit\'e, CNRS, Astroparticules et Cosmologie, F-75013 Paris, France \\
            \email{duverne@apc.in2p3.fr}
        \and 
            Aix Marseille Univ, CNRS/IN2P3, CPPM, Marseille, France 
        \and
            School of Physics and Astronomy, University of Minnesota, Minneapolis, Minnesota 55455, USA \\
             }

    \date{Received XXX; accepted XXX}

 
  \abstract
   {Core-collapse supernovae (CCSNe) are expected to produce intense bursts of neutrinos preceding the emergence of their electromagnetic counterparts. The prompt detection of such neutrino signals offers a unique opportunity to trigger early follow-up observations in the electromagnetic domain, enabling direct studies of the explosion mechanisms and progenitor properties.}
   {We aim to assess the feasibility and efficiency of an optical/near-infrared follow-up strategy for CCSNe discovered via neutrino bursts, by modelling the expected spatial distribution of events and simulating realistic observational campaigns taking into account the size of the localization's error box generated by triangulating the neutrino signals detected by multiple neutrino detectors .}   
   {We modelled the Galactic distribution of CCSNe, including the effects of interstellar extinction, and considered three main progenitor types: Wolf–Rayet stars, red supergiants, and blue supergiants. We also included the shock breakout phase in the electromagnetic signatures that could be detected following the neutrino burst. A simulated population of CCSNe was generated and detected by different networks of neutrino observatories, including IceCube, KM3NeT/ARCA, Super-Kamiokande, Hyper-Kamiokande, and JUNO. For each sample, the resulting skymaps were used as input for the \texttt{GWEMOPT} scheduler to produce optimized follow-up plans with two optical facilities: the Vera C. Rubin Observatory (LSST) and the TAROT robotic telescopes.}
   {Both LSST-VRO and TAROT exhibit comparable detection efficiencies for the simulated CCSN population. However, the smaller TAROT network achieves similar success rates while requiring fewer pointings to cover the CCSN's localization regions provided by the neutrino signal.}
   {Our simulations demonstrate that neutrino-triggered follow-up campaigns can effectively identify CCSN optical counterparts using both large and small facilities. Depending on the neutrino network configuration, the median number of pointings for the two tested optical facilities is of the order of 20 to 100 to find the EM emission. In addition, the number of images is larger for LSST than for smaller, large-field-of-view facilities by a factor of 2 to 4. The combination of rapid localization from neutrino detector networks and flexible optical scheduling significantly enhances the prospects for multi-messenger studies of nearby stellar explosions.}


   \maketitle
%

\section{Introduction}
    The core-collapse supernova SN~1987A currently constitutes the only confident joint observation of an SN in both EM and neutrino regimes. The explosion of the progenitor was primarily detected by three neutrino detectors: Irvine-Michigan-Brookhaven (IBM) water Cherenkov detector in the US, Kamiokande in Japan and Baksan Neutrino Observatory (BNO) in Russia \citep{1987PhRvL..58.1490H,1988PhRvD..38..448H}, about three hours before the first optical observations of the supernova (SN). Based on this experience, the SNEWS alert system \citep{2008AN....329..337S, 2024JInst..19P0017K} was designed to publicly distribute alerts in case a neutrino burst consistent with a SN emission was simultaneously detected by the currently active MeV-neutrino instruments. 
    In addition, the operations of optical, high-cadence surveys such as ATLAS \citep{2018PASP..130f4505T}, ZTF \citep{2019PASP..131a8002B}, or ASAS-SN \citep{2017PASP..129j4502K} in recent years and the upcoming Large Synoptic Survey Telescope-Vera Rubin Observatory (LSST-VRO) \citep{2009arXiv0912.0201L} led to a significant rise in the detection of optical transients, including SNe. In addition, the multi-messenger observations of GW170817 \citep{grb_2017, Goldstein_2017, mma170817, D_Avanzo_2018, Alexander_2017, Hallinan_2017, Troja_2017, Soares_Santos_2017} showed the potential of worldwide coordinated observations of astrophysical events. It led to the development of various optical networks of telescopes, such as GOTO \citep{GOTO} or GRANDMA \citep{mamie1, mamie2}. Consequently, time-domain astronomy has undergone significant developments and discoveries. These new optical observing capabilities can also be conjugated with the current and upcoming neutrino detectors—e.g., Ice-Cube\citep{icecube}, KM3NeT \citep{2024EPJC...84..885K,2019arXiv190602704K, 2016JPhG...43h4001A}, JUNO \citep{juno}, and Hyper-Kamiokande \citep{hyperk}—to enhance the probability of achieving multi-messenger observations of the next Galactic CCSN, as occurred in 1987 with the detection of SN1987A \citep{ivanez2023sn1987a,1987PhRvL..58.1490H,1988PhRvD..38..448H}. In this regard, there are also some systematic follow-ups of neutrinos detected by Ice-Cube in both the optical domain with ZTF \citep{2023MNRAS.521.5046S,2025arXiv250309426L} and the gravitational wave domain \cite{2023ApJ...959...96A}, although they are related to high-energy neutrino candidates that are outside the MeV energy range relevant for supernova neutrino bursts. Such a multi-messenger observation would be crucial for addressing various questions, such as the cooling mechanism underlying the ejecta cooling, estimating the time delay between the shock-breakout \citep{adams2013observing} and the neutrino emission. Each messenger brings information that is not accessible to the other, providing a more thorough understanding of the event. The MeV neutrino burst provides a snapshot of the progenitor's core during the collapse and allows to probe the internal structure of the progenitor \citep{2013ApJ...778...81K}. Moreover, neutrinos, being the first messengers to be emitted during the collapse, drive the follow-up observations. On the other hand, the EM emissions inform about the external regions such as the ejecta (mass, velocity and energy), in addition, spectral observations allow to probe the chemical composition of the progenitor.


    The current capabilities for detecting neutrinos emitted by a CCSN limit the horizon for a photons/neutrino joint observation to the Milky Way and its close-by companions, such as the Large Magellanic Cloud. Considering the CCSNe rate (about once per century) at the Galactic level, this type of event has to be considered rare. However, dedicated tools must be prepared to enable the follow-up observations to maximise the chances of a multi-messenger detection and the scientific return of a joint observation. These goals led to the creation of the SNEWS collaboration \citep{2008AN....329..337S, 2024JInst..19P0017K}, which enables the sharing of alerts between different neutrino observatories and the distribution of alerts to other communities to facilitate electromagnetic (EM) follow-up observations. However, the expected areas of localisation contours for a neutrino burst emitted by a CCSNe \citep{coleiro2020combining} are at the hundreds of square degrees level, which imposes strong constraints on the follow-up and requires specific strategies to enable detection of the EM counterpart. Such large localisations require important observational resources to maximise the chances of detecting the EM counterpart rapidly. Despite having its own challenges, it is noticeable that the follow-up of gravitational wave (GW) alerts has strong similarities with the follow-up of a neutrino event: the size of the localisation area, the need for rapid follow-up to discover the early EM counterpart, such as the shock breakout, and the rarity of events. Consequently, it is worth considering some of the lessons and methods provided by the GW follow-up in the context of neutrino astronomy. In particular, the lack of coordination between observatories and collaborations for the GW follow-up has been claimed to be responsible for the lack of EM counterpart detections during the O3 and O4 observing campaigns of LIGO, Virgo, and KAGRA \citep{2024arXiv240517558K}. To avoid this situation in the case of a Galactic CCSN, we propose a study that simulates a population of CCSNe emitting a neutrino burst observed in a network of detectors, composed of a combination of IceCube, Hyper-Kamiokande, KM3NeT-ARCA, and JUNO. The multiple detection then allows triangulating the source and producing a localisation map. The latter is then used as an input for a realistic follow-up campaign with the LSST-VRO and TAROT, a network of small, robotic facilities with large fields of view, that routinely participate in the GW follow-up. The first section describes the simulation setup used to generate the SNe population, the neutrino triangulation procedure, the optical lightcurves, and the planning of the follow-up campaign for the two envisioned follow-up configurations. The third section describes the results of the simulations and the detections of the CCSN in the optical and neutrino regimes. Eventually, we discuss the implications for the planning of the follow-up campaigns for LSST-VRO and the smaller facilities for the next neutrino-bright Galactic CCSN.

\section{Simulation setup}  
    In this section, we overview the process we use to simulate a population of Galactic supernovae, how we assess the detection of the neutrino burst and its localization and finally, the planning of the follow-up observations to find the EM emission with ground-based optical facilities. \\ \noindent
    We start by distributing a population of 500 sources throughout the Milky Way (MW), with three types of progenitors: red supergiants, blue supergiants and Wolf-Rayet stars.
    Then, we assign an optical lightcurve to each SN based on a template model, and we also include a simplified shock breakout (SBO) phase to estimate the chance of detecting it. We associate a sky localization probability density with the neutrino burst, referred to as a \textit{skymap} thereafter, by triangulating the location of the CCSN source. After estimating the localization, we plan the optical follow-up campaign with different optical facilities using the widely used \gwo algorithm for tiling the sky localization \citep{2019MNRAS.489.5775C}. First, we consider the LSST-VRO searching the neutrino skymap for the EM counterpart. Then we use TAROT, a network of smaller robotic facilities, and compare the respective performances in finding the EM emission of the SNe. 
    
    \subsection{Core-Collapse supernova population}
    \label{sec:pop}
    \subsubsection{Spatial distribution}
    \label{sec:space-distrib}
    The population of supernovae we use is distributed in the MW following the Monte-Carlo approach of \cite{adams2013observing}. The population is generated following a double exponential distribution:
    
    \begin{equation}
        \rho(R,z) = A e^{-\frac{R}{R_d}} e^{-\frac{|z|}{H}}
        \label{eq:space_distr}
    \end{equation}
    
    where \(R\) is the Galactocentric radius, \(z\) is the height above the galactic mid-plane and $H$ and $R_d$ are model scale lengths. This modelling assumes that the star formation follows the dust distribution in the MW \cite{adams2013observing}. Consequently, the scale length of the thin disk is used as the scale length of the dust distribution. Following the parametrisation of \cite{adams2013observing}, we use a scale length of $R_d=2.9$ kpc and $H = 95$ pc, based on the input parameters of TRIdimensional modeL of thE GALaxy (TRILEGAL), a population synthesis code for simulating stellar populations along any direction through the Galaxy \citep{girardi2012trilegal}. In this configuration, the Sun is placed at $H_{\odot} = 24$ pc above the disk's mid-plane at a Galactocentric radius of $R_{\odot} = 8.7$ kpc.

    \subsubsection{Progenitors distribution}
    We consider three distinct types of stellar progenitors: red supergiants (RSG), blue supergiants (BSG), and Wolf–Rayet stars (WR). We estimate the fraction of each progenitor type in the simulated population using available observational data. Hence, we use the ratio between BSG and RSG progenitors and the RSG to WR ratio to estimate the respective fractions. \\
    Using a survey of young star clusters, \cite{eggenberger} finds the following dependence for the BSG to RSG ratio:
    \begin{equation}
        \frac{B/R}{B/R_{\odot}} = 0.05 e^{\frac{3 Z}{Z_{\odot}}},
        \label{eq:btor}
    \end{equation}
    where Z is the metallicity, $Z_{\odot}=0.02$ is the metallicity of the Sun and $B/R_{\odot} \cong 3.0$ is the value of B/R at $Z_{\odot}$ \cite{eggenberger}. The study also defines two regions in the MW with significantly different $Z$: the Galactic centre (GC), with $Z=0.023$, and the galactic anti-centre (GAC), where $Z=0.016$ (see Tab. 3 of  \citealt{eggenberger}). The limit between the two is at a Galactocentric radius of 8.5 kpc. Consequently, based on Eq.~\ref{eq:btor}, we use for the SN population:
    \begin{equation}
        \begin{cases}
            B/R = 4.73, ~\textrm{for}~R \leq 8.5~\textrm{kpc} \\
            B/R = 1.65, ~\textrm{for}~R > 8.5~\textrm{kpc}.
        \end{cases}
    \end{equation}
    
    For the RSG to WR ratio, denoted $R/W$, we use the work presented in \cite{massey}, where WR and RSG candidates are identified in several galaxies in the MW surroundings with various techniques (spectroscopy, photometry and image subtraction). $R/W$ shows a flat behaviour with the metallicity in the survey for $Z\geq0.005$, and \cite{massey} shows that it is consistent with several evolutionary models. Consequently, to generate the progenitor population, we use the R/W values tabulated in Tab. 3 of \cite{massey}. The latter uses three luminosity cuts to count the number of RSG in the surveyed galaxies. For this work, we use the luminosity cut of log $L/L_\odot \geq 4.8$, as this corresponds to progenitors with a mass of $\sim$15$M_\odot$ that corresponds to the typical mass of the observed progenitor in the Local Group \citep{smartt}. However, the Small Magellanic Cloud (SMC) has a metallicity of $Z=0.003$, which is below the $Z\geq0.005$ limit where the $R/W$ is flat and returns a $R/W$ that is significantly larger than the rest of the data. In addition, the SMC metallicity is significantly smaller than the MW metallicity provided in \cite{eggenberger}. Eventually, we obtain $R/W = 2.2$ by computing the median of the metallicities tabulated in Tab. 3 of \cite{massey}, without including the SMC data point. 
    
    Eventually, we denote $b$ the fraction of BSG, $r$ the fraction of RSG, $w$ the fraction of WR stars, $\alpha$ the B/R ratio, and $\beta$ the R/W ratio. Considering that $b+r+w=1$, we have:
    \begin{equation}
            \begin{cases}
            b = \frac{\alpha}{1 + \alpha + 1/\beta}, \\
            r = \frac{1}{1 + \alpha + 1/\beta}, \\
            w = \frac{1/ \alpha}{1 + \alpha + 1/\beta}.
            \end{cases}
    \end{equation} 

    The $b$, $r$ and $w$ values are tabulated in Tab.~\ref{tab:ratios}, along with the numbers of each progenitor type in the fourth row. 
    
        \begin{table}[h!]
        \caption{
            Fraction of the different progenitors for the SN population in the GC and GAC region of the MW. The last row provides the numbers for each type of progenitor in the population.
           } 
        \begin{center}
        \begin{tabular}{c|c|c|c}
        \hline
        \hline
        Fraction  & $r$ & $b$ & $w$  \\
        \hline
        GC  & 0.16  & 0.76 & 0.074 \\
        GAC & 0.32  & 0.53 & 0.15 \\
        \hline 
        \hline
        Number of progenitors & 342 & 107 & 51 \\
        \hline
        \end{tabular}
        \label{tab:ratios}
        \end{center}
        \end{table}

    \subsection{Neutrino burst triangulation}
    \label{sec:network}
    In the case of a galactic supernova observation, the neutrinos are expected to be detected before the EM emission. Therefore, this is the messenger that will provide the earliest pieces of information, particularly the estimation of the source's position. Based on the last multi-messenger event, GW170817, the rapid localization plays a key role in the EM follow-up observations \citep{grb_2017}. Accordingly, this places the neutrinos' localization in a central position for guiding the EM observations of the next Galactic CCSN.\\ \noindent    
    For this work, we assume that the neutrino signal is emitted as a burst and that it is detected by at least three detectors, allowing us to triangulate the source's position following the \lcm method proposed in \citep{coleiro2020combining}. We consider five detectors for the localization: Hyper-KAMIOKANDE (HK) \citep{hyperk}, Super-Kamiokande (SK) \citep{FUKUDA2003418}, Ice-Cube (IC) \citep{icecube}, KM3NeT-ARCA \citep{2019arXiv190602704K, 2016JPhG...43h4001A} and JUNO \citep{juno}. However, as SK is expected to be replaced by HK in the next decade, we are not considering a five-instrument configuration. Currently, only SK, IC, and KM3NeT are active, and this is the first configuration we are simulating, even though KM3NeT is not yet fully deployed. We are considering two other three-instrument configurations: IC-ARCA-HK and IC-ARCA-JUNO, assuming that SK has been decommissioned and that either JUNO or HK has replaced it, depending on whichever starts observing first. Eventually, we also use a best-case configuration with IC-ARCA-HK-JUNO, where four instruments detect the neutrino burst. For each configuration, we generate the skymaps for the whole CCSN population described in Sec.~\ref{sec:pop}.

    The \lcm method presented in \cite{coleiro2020combining} compares neutrino light curves measured by different detectors to determine their relative arrival-time delays. The light curves are modeled using a simplified neutrino emission model and a detector response based on effective volumes and background rates, and the comparison is performed using cross-correlation or chi-square minimization methods. The supernova position is then reconstructed by triangulating the delays between multiple detectors. Skymaps are then generated with a HEALPIX histogram \citep{gorski2005healpix,singer2022healpix} by computing expected time-of-arrival delays for each sky position and evaluating a $\chi^2$ statistic for every detector pair. Summing these contributions yields a total $\chi^2$ , which is converted into a probability map after identifying the minimum $\chi^2$ . By repeatedly sampling detector delays within their uncertainties, a final probability distribution over the sky is obtained. This results in a skymap indicating the most likely CCSN location. This method is implemented using an adapted version of the authors’ public code\citep{coleiro_2020_3779943} that allows to create skymaps that are compatible with \gwo, the algorithm used to make the follow-up campaign (see Sec.~\ref{sec:plan}.

    The skymaps used for this work are made with 64 pixels per side, which corresponds to approximately 0.84 square degrees per pixel, and by sampling the $T_{0,{ij}}^\mathrm{match}$ distribution depending on the neutrino network configuration. For the configurations that include HK, we use 30,000 samples; for the others, we use 100,000 samples. These choices allow us to keep the skymap generation time at the $\sim$minutes level. In addition, the pixel size of the skymap is smaller than the field of view of the optical instruments used for the follow-up campaign, as shown in Tab.~\ref{tab:gwemopt_param}, which avoids missing any region of the sky because of the tessellation. The network configurations with HK require fewer iterations to converge because the detector is more sensitive in the energy range of the neutrinos emitted during a CCSN (see e.g. \cite{coleiro2020combining} and references therein). 


    \subsection{ccSNe Lightcurve models}
    \label{sec:lc}
\begin{figure}[h!]
        \includegraphics[width=\columnwidth]{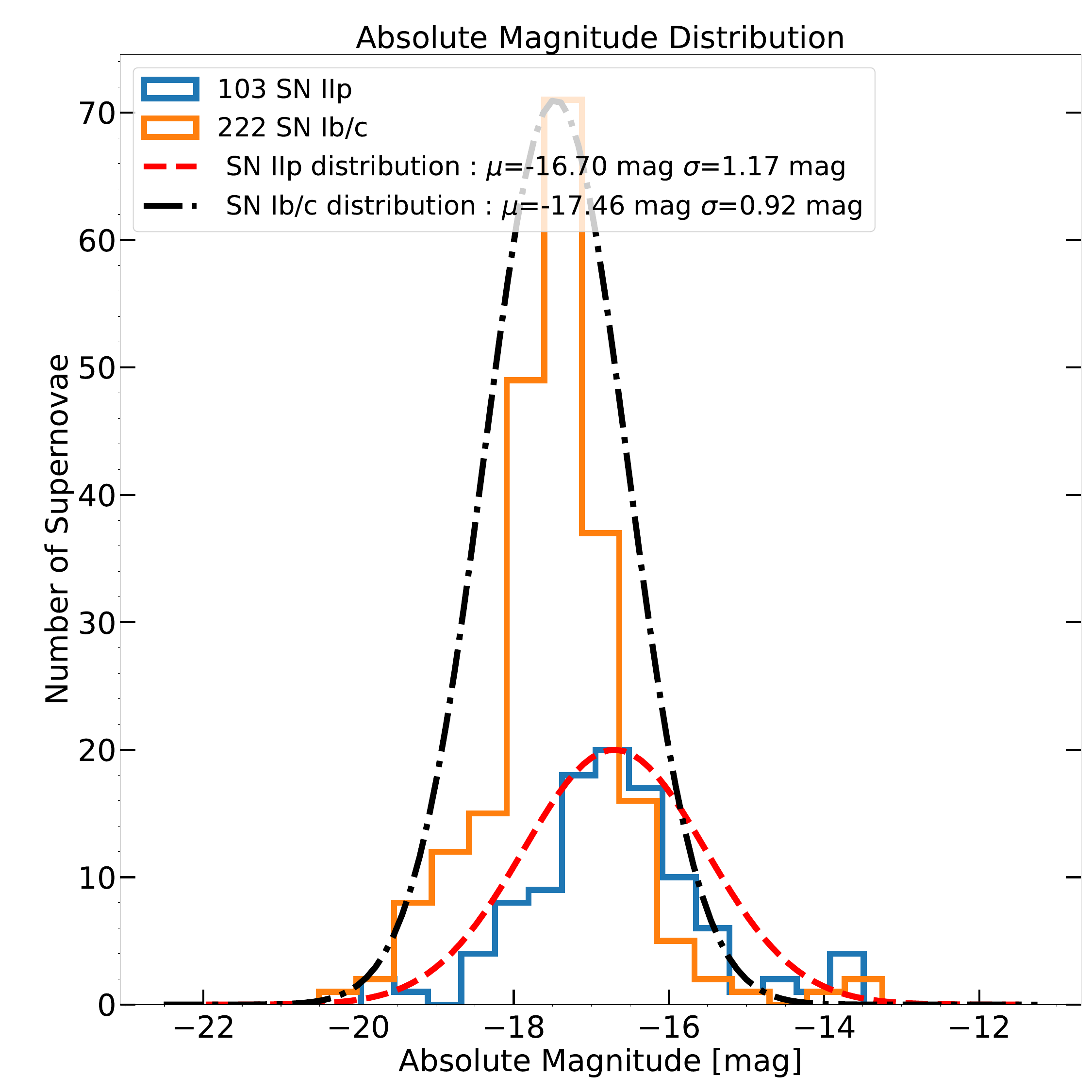}
        \caption{Absolute magnitude distribution of 103 observed type IIp SNe and 222 type Ib/c observed SNe. The data are taken from \cite{li2011nearby} and \cite{perley2020zwicky} for the SN IIp and only the latter for the SN Ib/c. The red curve shows the Normal distribution with mean $\mu = -16.70$ mag and standard deviation $\sigma = 1.17$ mag used to generate the SN IIp population. The black curve shows the Normal distribution with mean $\mu = -17.46$ mag and standard deviation $\sigma = 0.92$ mag used to generate the SN Ib/c population.}
        \label{fig:abs-mag}
    \end{figure}

    For each CCSN injected following the procedure in Sec.~\ref{sec:space-distrib}, we associate a lightcurve in optical/NIR bands to estimate the magnitude at the detection in Sec.\ref{sec:plan}. For the RSG and BSG progenitors, we use a type IIp SNe template based on the models presented in \cite{1999ApJ...521...30G}. This choice is motivated by the fact that the majority (about 60\%) of all the observed CCSN in the Local Group are of this type \cite{smartt}. On the other hand, WR stars are expected to lead to type Ib/c SNe \citep{2013ApJ...778...81K}, which represent 29\% of the observed SN \cite{smartt}. Consequently, we use a template lightcurve based on \cite{2005ApJ...624..880L} to estimate the SNe magnitude. The lightcurves for SN IIp and SN Ib/c differ mostly in the early phases of the explosion, when the latter are slightly brighter by $\sim$2 magnitudes during the first hours, as illustrated in Fig.~\ref{fig:lc-ex}. \\ \noindent
    The absolute magnitude of the individual SN IIp of the population is generated following a Normal distribution with mean $\mu = -16.70$ mag and standard deviation $\sigma = 1.17$ mag. For the SN Ib/c SNe, we used a Normal distribution with  $\mu = -17.46$ mag and standard deviation $\sigma = 0.92$ mag. These values are estimated based on publicly available observational data of SNe in recent optical surveys. We used the data provided in \cite{perley2020zwicky} for the SN Ib/c and in \cite{li2011nearby} and \cite{perley2020zwicky} for the SN IIp. We show the distributions used to generate the absolute magnitudes for the observed SNe in Fig.~\ref{fig:abs-mag}, where the black curve corresponds to the SN Ib/c and the red curve to the SN IIp.

    \subsection{Dust and Extinction}
    \label{sec:dust}
\begin{figure*}
        \includegraphics[width=\textwidth]{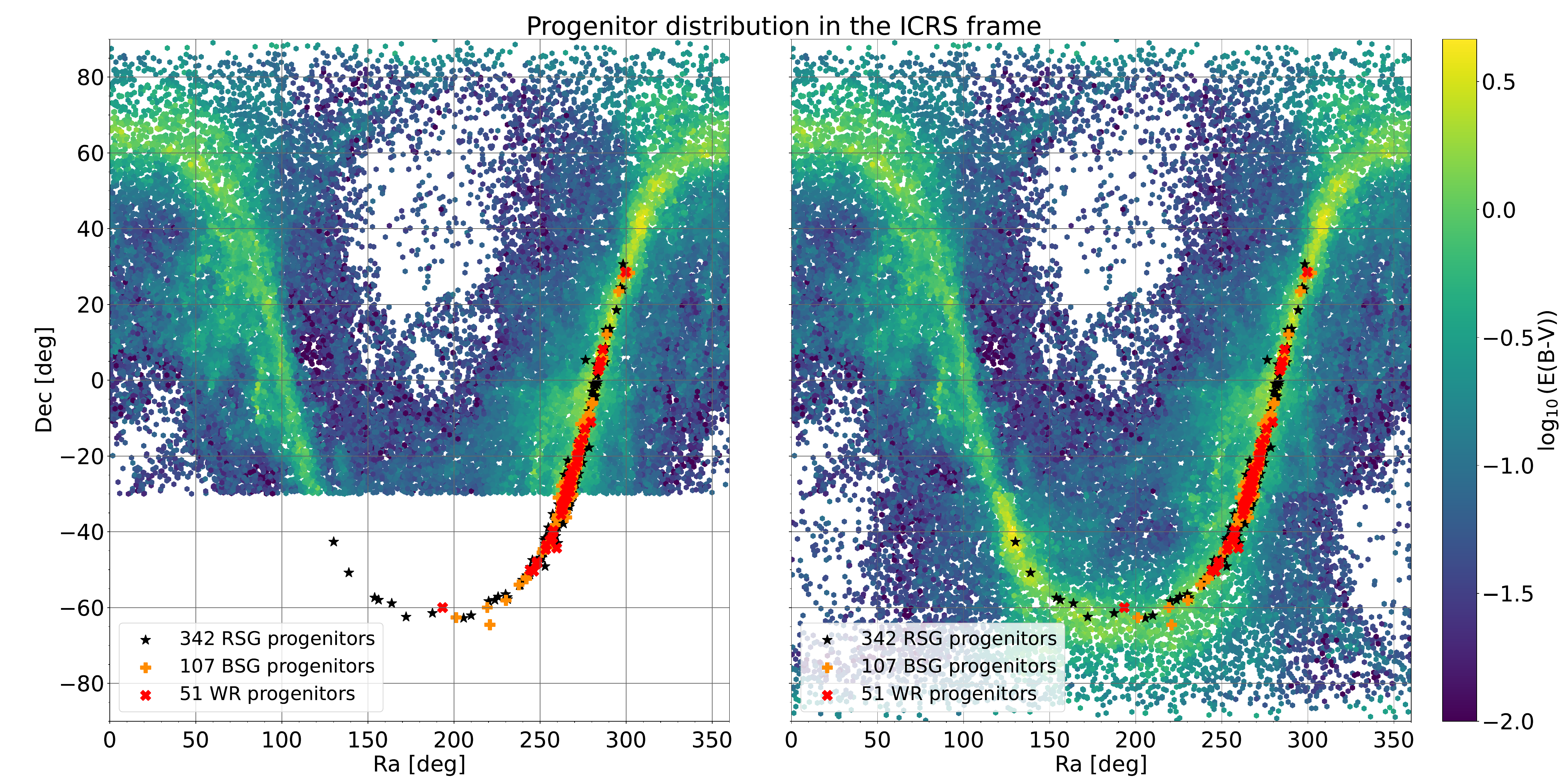}
        \caption{Distributions of the E(B-V) colour excess for a sample of 50000 positions in the MW using the \texttt{BAYESTAR} dust map that were generated following the same distribution as in Eq.~\ref{eq:space_distr}. The left panel shows the colour excess results without the transformation presented in Eq. \ref{eq:ebv_transf}, and the right panel shows the results that include them. Both plots show the 500 injected SNe, with the RSG progenitor represented by black stars, the BSG by orange crosses, and the WR ones by red crosses.}
        \label{fig:icrs}
    \end{figure*}

\begin{figure}
        \includegraphics[width=\columnwidth]{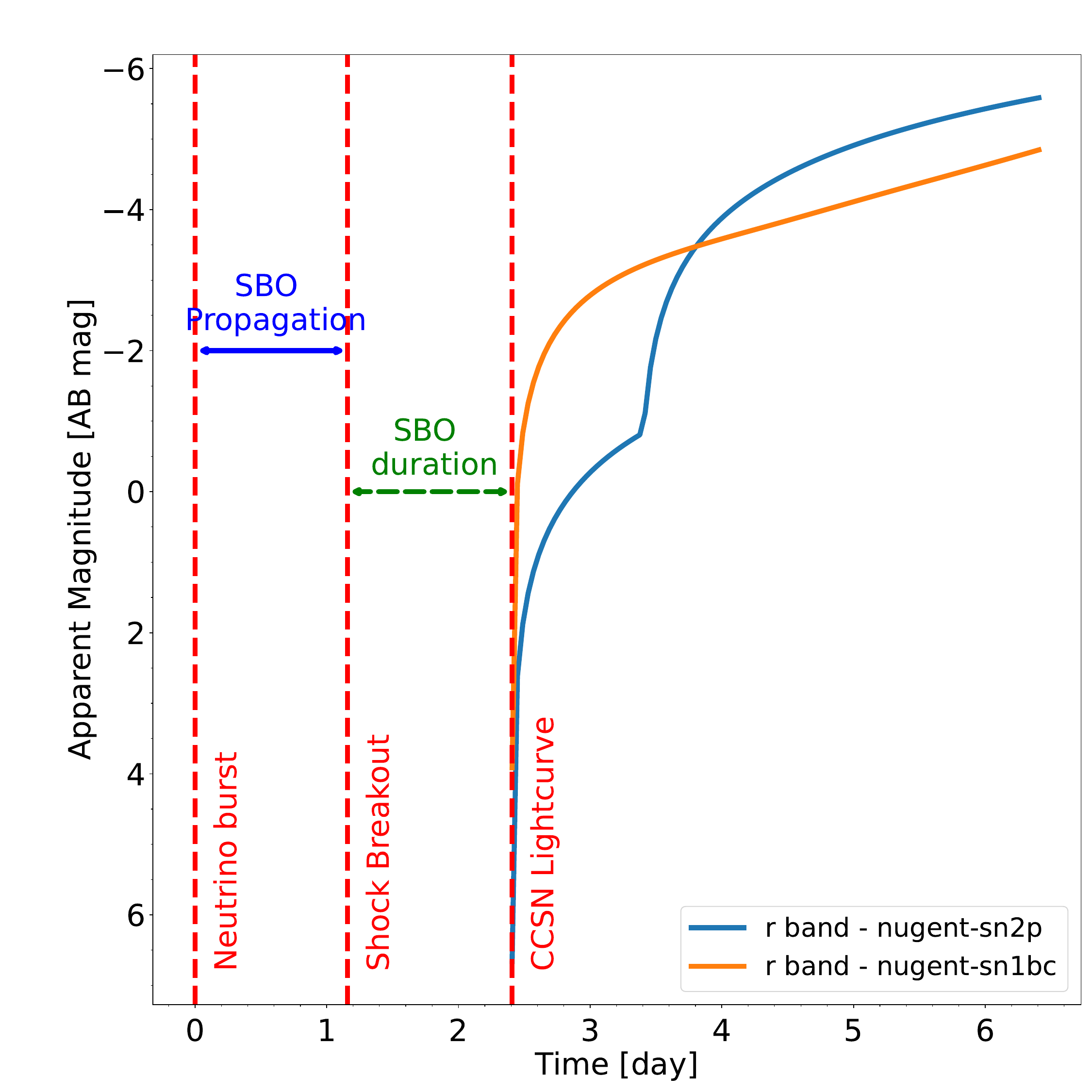}
        \caption{Temporal diagram with the events following the neutrino burst that triggers the optical observations. The duration of the shock breakout propagation is visible as the arrows in blue and the shock breakout duration is presented as the green arrow. However, the values presented in the figure are arbitrary and chosen to make the plot clearer. The orange lightcurve is based on the template presented in \cite{1999ApJ...521...30G} for the Type IIp events. The blue curve corresponds to the Type Ib/c CCSN and is based on the template presented in \cite{2005ApJ...624..880L}.}
        \label{fig:lc-ex}
    \end{figure}

    Dust absorption in the Milky Way significantly affects SNe's observed brightness and colours by scattering and absorbing light, increasing the SNe magnitude via extinction. To quantify this effect for the SNe population, we estimate the colour excess parameter $E(B-V)$ using the 3D \texttt{BAYESTAR} dust map \citep{green20193d} (2019 version). The latter is built by the combination of the parallaxes provided by the Gaia satellite \citep{gaia2016gaia} and the photometric observations from the Pan-STARRS (PS) optical survey \citep{chambers2016pan} and the near infra-red (NIR) survey 2MASS \citep{skrutskie2006two}. Using the simulated Right Ascension (RA) and Declinations (Dec), we request the colour-excess E(B-V) term at the equatorial position of each sample of the population described in Sec.\ref{sec:pop}. However, the PS observations are limited to declinations above -30$^{\circ}$; consequently, \texttt{BAYESTAR} does not provide the colour excess term in these regions, as illustrated on the left panel of Fig.~\ref{fig:icrs}. As a significant part of the population of CCSN we simulated is located in this region, we apply the transformations presented in Eq.\ref{eq:ebv_transf} to the coordinates of the samples for which the \texttt{BAYESTAR} map has no data. They provide realistic $E(B-V)$ values for these injections under the hypothesis that the dust distribution is symmetric in the MW. 
        \begin{align}
        \begin{split}
        \begin{cases}
        \label{eq:ebv_transf}
          \textrm{RA'} = \textrm{RA} - \pi \\ 
          \textrm{Dec'} = - \textrm{Dec}.
        \end{cases}
        \end{split}
        \end{align}

    Eq.~\ref{eq:ebv_transf} allows us to estimate the colour excess terms in the Southern Hemisphere using a rotation from a CCSN at position (RA, Dec) to (RA', Dec'). The latter are the coordinates used to estimate $E(B-V)$ from \texttt{BAYESTAR}. The process is illustrated on the right panel of Fig.~\ref{fig:icrs}, where the regions below -30$^{\circ}$ are completed. The figure also includes the position of the 500 injected CCSN broken down by progenitors.
    
     Once the colour excess $E(B-V)$ value is estimated, we compute the extinction $A_V$ for each SN sample of the population by propagating the absorption along the line of sight. In that regard, we use:
    \begin{equation}
        A_V = \frac{R_V}{E(B-V)}, 
    \end{equation}
    where $A_V$ is the extinction in the $V$ band and $R_V=3.1$, following \citep{1989ApJ...345..245C} approach. This value is averaged over the whole galaxy. Even if some regions of the MW are known to have larger values of $R_V$ (e.g. dense molecular cloud for example \citealt{1993A&A...274..439J, 2010A&A...522A..84O}), they are not expected to have significant effects on the final results \citep{adams2013observing}. Eventually, we include the $E(B-V)$ colour-excess in the lightcurves described in Sec.~\ref{sec:lc} to have the extinction-corrected magnitude of the SN. The lighcurve models and the extinction correction are managed using the \texttt{sncosmo} Python package \citep{barbary2016sncosmo,barbary_2025_15019859}.  

    \subsection{Shock Breakout}
    The first observable EM counterpart expected to arrive after the neutrino emission is the shock breakout (SBO). This radiative precursor consists of a flash in the X-ray/UV domains emitted when the shock created during the collapse reaches and bursts through the surface of the progenitor star. This emission is rarely seen and a limited number of examples are known: GRB 060218/SN 2006aj \citep{2006Natur.442.1008C}, XRT 080109/SN 2008D \citep{2008Natur.453..469S}, and SNLS-04D2dc \citep{2008Sci...321..223S}. To increase the number of SBO detections, using other messengers such as neutrinos or a gravitational wave burst would help guide the observations by providing a starting point for the explosion and a rough estimation of the source's localization, as detailed in Sec.~\ref{sec:network}. A multi-messenger joint observation of neutrinos and an SBO would also allow constraining the shock propagation time and the properties of the progenitor stars \citep{2013ApJ...778...81K}. Therefore, conducting a dedicated follow-up campaign to identify the neutrino source would enhance the characterisation of the progenitor and the explosion. However, the delay between the neutrino burst and the SBO emission, along with the pulse duration, varies from star to star and strongly depends on the progenitor's type. For modelling the SBO and evaluating the number of detections, we use the polytrope model provided in \cite{2013ApJ...778...81K} and also used in \cite{adams2013observing}. It provides an estimate for the three types of progenitors covering a range of masses between 11 and 30M$_{\odot}$ for the RSG, 12-16M$_{\odot}$ for the BSG and 16-35M$_{\odot}$. These estimates are presented in Fig. 2 of \cite{2013ApJ...778...81K}, and we use them to implement the timing values for our population's different progenitors. As the observed SN IIp in the Local Group are mostly below 16M$_{\odot}$ \citep{smartt}, we use the value for the $\sim$15M$_{\odot}$ simulated cases. The implemented values are summarized in Tab.~\ref{tab:sbo}. 

    Although the SBO is expected to be mainly emitted in the X-ray/UV and then constitutes a target for space-based observatories, it also emits in the optical bands \citep{adams2013observing,2010ApJ...725..904N}. This opens the searches for ground-based facilities, and we use Fig. 3 of \cite{2010ApJ...725..904N} to estimate the absolute peak magnitudes of SBO in the optical bands as follows. Based on the current performance in GW and GRB follow-up with robotic instruments \citep{mamie2}, we estimate the delay between the neutrino alert, its processing to create the skymap and distribute it to the facilities to be $\sim$5 minutes. Based on Fig. 3 of \cite{2010ApJ...725..904N}, this yields an absolute magnitude of $\sim$-11 mag for BSG and $\sim$-13 mag for RSG. Considering the SBO duration (about 2s) and the delay between the neutrino burst and the SBO emission (about 50s) for WR stars \citep{adams2013observing}, we do not include them in the candidate for SBO observations with ground-based instruments. These results are also included in Tab.~\ref{tab:sbo}. For the SBO, we include the extinction using the \texttt{extinction} package \citep{barbary_2016_804967} and we use the $R_V=3.1$ \citep{1989ApJ...345..245C} value, as we do for the lightcurves presented in Sec.~\ref{sec:lc}.
        
    \begin{table*}
    \begin{center}
        \caption{
        Shock breakout characteristics for the three types of progenitors used in the population described in Sec.\ref{sec:pop} - Red Supergiant (RSG), Blue Supergiant (RBG) and Wolf-Rayet (WR) stars. The timing - delay and duration - are extracted from Fig. 2 of \cite{2013ApJ...778...81K}. The absolute magnitudes of the SBO are extracted from the Fig. 3 of \cite{2010ApJ...725..904N}. The very short delay between the neutrino burst and the SBO, combined with its short duration for WR progenitors, makes the SBO detection with a ground-based optical facility impossible. Hence, we do not include an absolute magnitude for the WR.
       } \label{tab:sbo}
    \begin{tabular}{c|c|c|c}
    \hline
    \hline
    Progenitor  & SBO delay & SBO duration & Absolute magnitude \\
         -      &    [s]    &      [s]     &        [mag]       \\
    \hline
    RSG & $1\times10^5$  & $3\times10^3$ & -13 \\
    BSG & $3\times10^3$  & $1\times10^2$ & -11 \\
    WR  &      50        &       2       &  -  \\
    \hline 
    \end{tabular}
    \end{center}
    \end{table*}

    \subsection{Optical Telescopes Configuration}
    \label{sec:tel-conf}
    We consider two complementary observational configurations to design a follow-up campaign to identify an EM counterpart following a neutrino alert emitted by a CCSN. The first employs the LSST-VRO, which combines a 9.6 deg$^2$ FoV and can reach a limiting magnitude of r$\sim$24.5 mag in a single 30s exposure \citep{2019ApJ...873..111I}. Moreover, the LSST-VRO pipelines include image-subtraction and transient identification algorithms, which increase the chance of rapidly detecting the counterpart and distributing the information via alert brokers \citep{Smith_2019,2021AJ....161..107M,2019A&A...631A.147N,2021AJ....161..242F,2021MNRAS.501.3272M}. For the simulated plans, we assume that LSST-VRO will follow the neutrino alerts as soon as possible, without any other constraints than the observability. However, due to its survey scheduling and sky visibility constraints, LSST-VRO cannot always provide immediate follow-up within the low-latency window critical for rapidly fading transients. 
     
    For a further illustration, we adopt a second configuration with telescopes for which the design is optimized for rapid follow-up, namely the TAROT network \citep{2008PASP..120.1298K,2019ApJ...886...73N}. The latter is a three-instrument network of small, robotic facilities with apertures ranging from 18 to 25 cm. The three instruments are distributed worldwide - one is in Chile (TCH), one in southern France (TCA), and one on Réunion Island (TRE). This allows covering the whole sky and having instruments in various time zones, making it easier to launch observations rapidly, as it increases the chance of having a facility ready to observe. Although these telescopes typically achieve shallower depths - r$\sim$16-18 in 180s and narrower fields of view compared to LSST-VRO, their automated operations, low-latency response (often within minutes of the alert), and global coverage enable prompt follow-up at any time of the night \citep{2006ApJ...638L..71B}. Furthermore, they are routinely used to work with \gwo plans \citep{mamie1,mamie2}. Combining these two configurations enables us to compare the rapid temporal coverage with the deep imaging capability provided by a large facility, thereby maximizing the probability of detecting and characterizing the EM counterpart.

    \subsection{Follow-up Campaign planning}
    \label{sec:plan}
    After the neutrino burst detection, the localization information has to be distributed to external communities to enable the setup of a coordinated campaign to maximize the chances of early detection of EM counterpart(s). This approach has been proven successful for the GW170817 event \citep{grb_2017, Goldstein_2017, mma170817, D_Avanzo_2018, Alexander_2017, Hallinan_2017, Troja_2017, Soares_Santos_2017} and has led to the detection of counterparts across the whole EM spectrum. Since 2017, the strategies and tools for organizing such campaigns have been developed and are now mature. Therefore, we propose to adapt them to the current situation for searching for an EM counterpart to a neutrino-bright CCSN. As presented in \cite{coleiro2020combining}, the size of the localization areas of the initial neutrino burst is at the 100-1000 deg$^2$, and these values are consistent with the results presented in Fig.~\ref{fig:cra}. These values are similar to those provided by the GW interferometers \citep{2023PhRvX..13d1039A,mamie1,mamie2}, which opens up two strategies for finding the neutrino source. One best suited to narrow field of view (FoV) optical instruments consists of establishing a list of potential progenitors compatible with the sky localization and observing them sequentially to find the source. Such a list of CCSN candidates has been proposed in \cite{2024MNRAS.529.3630H}. This approach can be compared to the \emph{galaxy-targeting strategy} used for GW follow-ups; see, e.g., \cite{2020MNRAS.492.4768D,2016ApJ...820..136G} and references therein for more details. On the other hand, the second option is to use wide FoV facilities to tile the credible region provided by the burst triangulation. This method for blind search of EM counterpart has been widely used in the previous years in the follow-up of GW events by various groups \citep{mamie1,mamie2,2024PASP..136k4201A,2019ApJ...881L...7G,2019PASP..131d8001C,2017ApJ...848L..33A}. The tool that is usually used for making the observation plan is Gravitational-Wave ElectroMagnetic OPTimization (\gwo) \citep{2018MNRAS.478..692C}. This software is designed to optimize the follow-up of GW events with EM ground-based observatories. \gwo plays a key role for the multi-messenger astronomy community, as it helps coordinate observations between different telescopes following a GW detection. Considering that the skymaps in this work are similar to the GW localizations, we adopt this tool to simulate realistic follow-up observation campaigns of neutrino detections. 
    
    \gwo takes the skymaps produced in Sec.~\ref{sec:network} and starts by tessellating the sky into tiles corresponding to the FoV of the instrument used for the follow-up, ensuring minimal overlap. Standardized configuration files are used to manage the various characteristics of the instruments, and the parameters for the instruments considered in this work are listed in Tab.~\ref {tab:gwemopt_param}. Then the algorithm determines the best fields to observe. For this work, we use the \emph{greedy} tilling algorithm based on the emcee process \cite{2013PASP..125..306F,2019MNRAS.489.5775C}. The tiles are placed simultaneously to optimize the observations. The algorithm selects the tiles of the sky's tessellation associated with the optical instrument covering the highest probability. Then, the tiles are ranked by the order of their integrated probability. The time allocated to each tile can also be optimized, however, as we are simulating bright transients, we use a fixed observing time for each tile. For LSST-VRO, we use 30s exposures, which allow us to reach $r\sim$24.5 mag (see \citealt{2019MNRAS.489.5775C} and Tab.~\ref{tab:gwemopt_param}. For the TAROT instruments, we use 180s exposures to reach $\sim$18 mag in TCA and TCH, and $\sim$ 16 mag in TRE (see \citealt{mamie2} and Tab.~\ref{tab:gwemopt_param}). We only consider observations in one filter to avoid filter changes during the blind search phase. We use the $r$ band as it is usually the default filter used during transient follow-up, as it provides the best sensitivity in the optical domain. Once the tilling has been made and the telescope time estimated, \gwo schedule the pointings depending on the observational constraints (e.g. the necessity for revisits during the night, airmass, etc...). We use the \emph{greedy} scheduling algorithm proposed in \gwo to keep the follow-up simple. In that configuration, the tiles with the highest probability in a given time segment are observed first \cite{2019MNRAS.489.5775C}. We eventually end up with an ordered list of tiles, each one corresponding to a position the telescope will observe.
    
        
        \begin{table*}
        \begin{center}            
        \caption{
            Configuration used in \gwo for the instruments used for this work. The LSST-VRO parameters are extracted from \cite{2019MNRAS.489.5775C} and the TAROT instruments information are taken from \cite{mamie2}
           } 
        \begin{tabular}{ccccccccc}
            \hline
            \hline
        Telescope &  Latitude  & Longitude & Elevation &       FOV          & Band & Exp. time & U.L.  \\
            -     &    [deg]   &   [deg]   &    [m]    &       deg          &   -  & [s]       & [mag] \\
        \hline
        LSST-VRO  & -30.1716   & -70.8009  &  2207.0   &      1.75          &   r  &    30     & 24.4  \\
        TAROT/TRE & -21.201387 & 55.407463 &  970.0    &  4.2 $\times$ 4.2  &   r  &   180     & 16    \\
        TAROT/TCA & 43.75203   & 6.92353   &  1320.0   & 1.85 $\times$ 1.85 &   r  &   180     & 18    \\
        TAROT/TCH & 29.2608    & -70.7322  &  2347.0   & 1.85 $\times$ 1.85 &   r  &   180     & 18    \\
        \hline
        \end{tabular}
        \label{tab:gwemopt_param}
        \end{center}
        \end{table*}
    
    \subsection{Randomizing the CCSN position}
    \label{sec:rando}
\begin{figure*}
        \includegraphics[width=\textwidth]{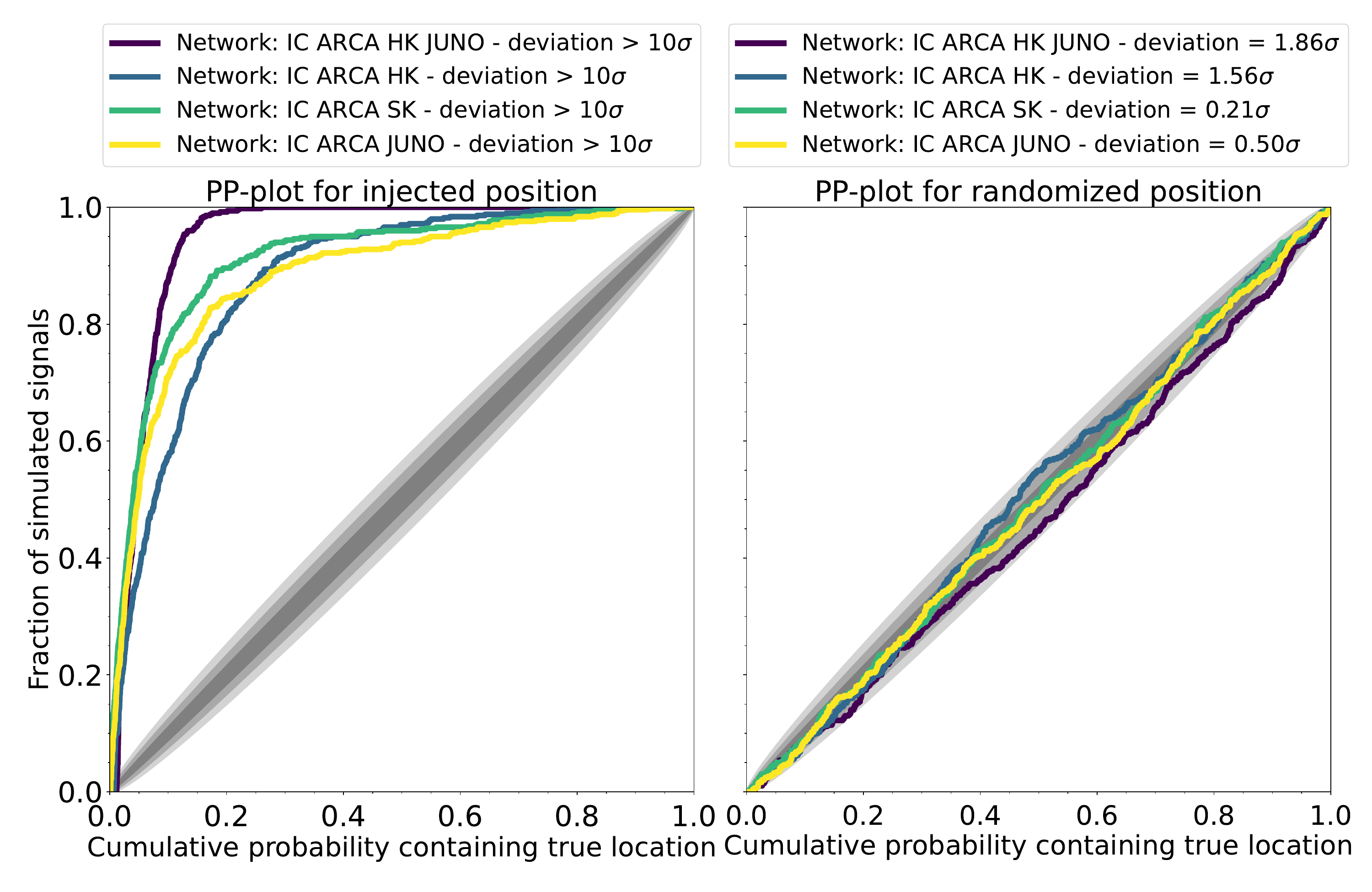}
        \caption{Percentile-Percentile plot for the four neutrino network configurations. The left and right panels show the results, respectively, before and after the randomization of the SNe position. For both panels, the dark blue curve corresponds the network IC-ARCA-HK-JUNO, the blue curve corresponds the IC-ARCA-HK network, the green curve corresponds the IC-ARCA-SK network, and the yellow curve corresponds to the IC-ARCA-JUNO network. The grey area corresponds to the 1, 2, and 3 $\sigma$ deviations from the diagonal. On the left panel, the curves are largely deviating from the diagonal by more than 10$\sigma$. On the right panel, all the curves are diagonal and compatible with a uniform \emph{search probability} distribution. The deviations for each network configuration are indicated in the legend of the plot.}
        \label{fig:pp-plot}
    \end{figure*}

    After estimating the position of the CCSN using the \lcm algorithm following the method described in Sec.~\ref{sec:network}, we introduce a randomization step to avoid systematically having a true position at the center of the skymap, in the highest probability region. As the planning of the follow-up usually starts in this region of the skymap, it would eventually lead to an underestimation of the time required to find the EM counterpart to the neutrino emission. The new position of the CCSN is chosen by picking one of the pixels following the probability distribution produced by \lcm. This ensures that the randomization remains consistent with the underlying spatial probability distribution while preserving the angular resolution defined by the HEALPix grid. This new position is eventually used as \emph{true} position for the following sections. \\ \noindent
    To validate this approach, we employ a test based on a \emph{percentile-percentile} plot (PP-plot). We compute the cumulative sum of the probability contained in the skymap pixels, starting from the highest-probability pixel and proceeding in order of descending probability, and stopping at the pixel that contains the true position. This sum corresponds to the cumulative probability in the smallest area that contains the true location, and is referred to as the \emph{search probability} in the following. Repeating this procedure for the whole population of simulated CCSN, we then make a cumulative histogram of the \emph{search probability}. This histogram is called a PP-plot and allows one to evaluate whether the localization is self-consistent by comparing the \emph{search probability} distribution to a uniform distribution. In the case of a consistent localization, the histogram is expected to be diagonal, indicating that the search probability distribution is uniform. In other words, one should find $p$\% of the simulated signals to have their true location within the $p$\% credible region. This procedure is performed for the three network configurations described in Sec.~\ref{sec:network}, using both the injected position and the randomized one as the \emph{true} position to estimate the \emph{search probability}. Fig.~\ref{fig:pp-plot} shows the PP-plots for the various networks. On the left panel, the curves show a clear deviation from the diagonal, interpreted as an overestimation of localization uncertainties due to the localization method. The uncertainties are constructed around the injected position, where the highest probability pixels are located. On the contrary, after the randomization of the position, the PP-plots presented in the right panel of Fig.~\ref{fig:pp-plot} show no more visible deviation within a 3-$\sigma$ interval. Consequently, after the randomization, the localizations become self-consistent and the process is considered valid.

\section{Results}
    \label{sec:results}
    After having described the details of how the simulations are performed, in this section we explore and discuss the results for the different choices of neutrino configurations and follow-up campaign.
    
    \subsection{Neutrino Localisation}
    \label{sec:nu-loc}
\begin{figure}[h!]
        \includegraphics[width=\columnwidth]{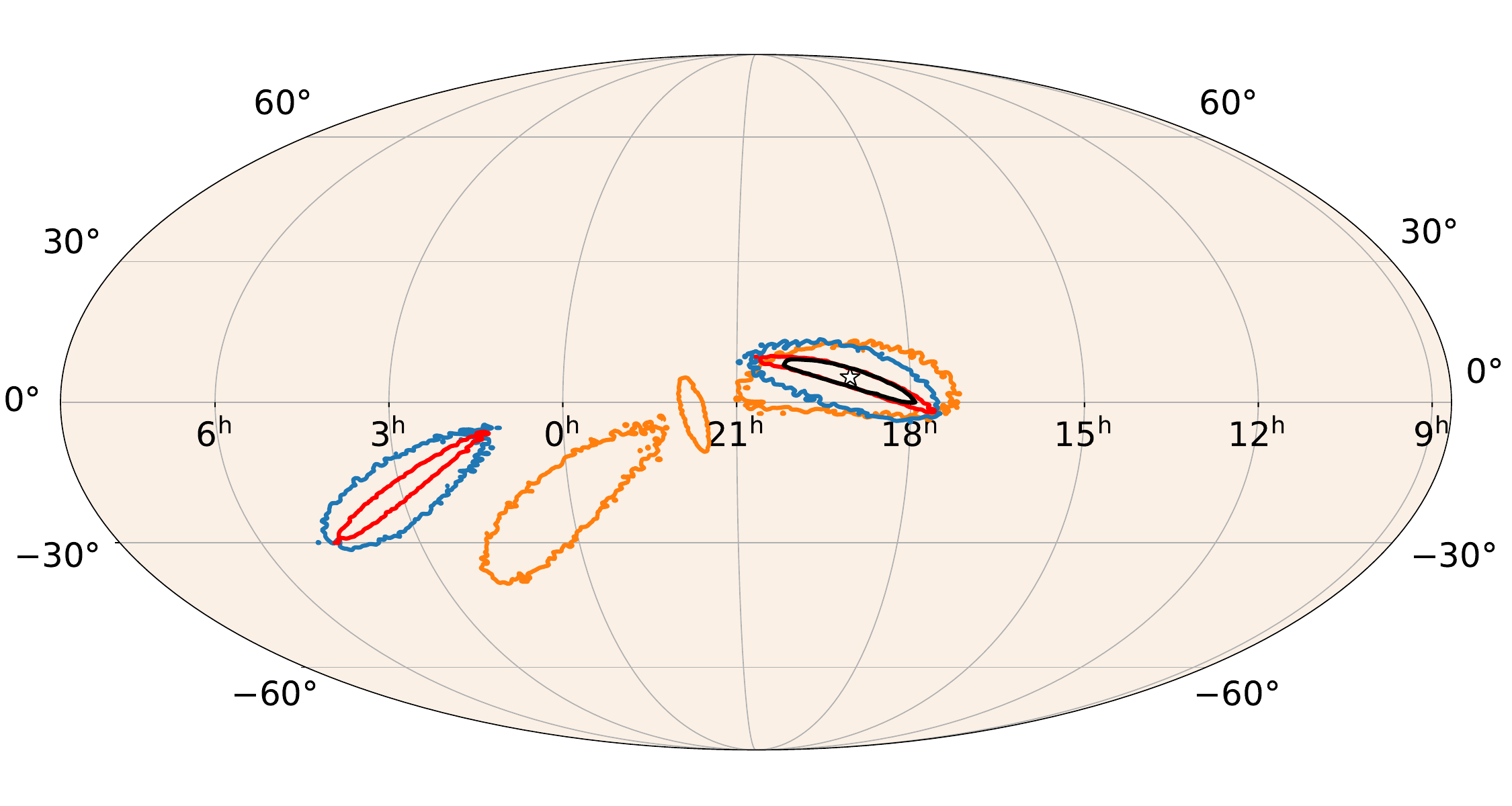}
        \caption{Skymaps obtained with \lcm for injection n$^{\circ}$69 for the different neutrinos network configurations described in Sec.~\ref{sec:network}. The black contour corresponds to the IC-ARCA-HK-JUNO configuration, the blue contour corresponds to the IC-ARCA-SK configuration, the orange contour corresponds to the IC-ARCA-JUNO configuration and the red contour corresponds to the IC-ARCA-HK configuration. The plot is made in the ICRS frame. The position of the CCSN injection is shown as a star marker.}
        \label{fig:skm-ex}
    \end{figure}

\begin{figure*}
        \includegraphics[width=\textwidth]{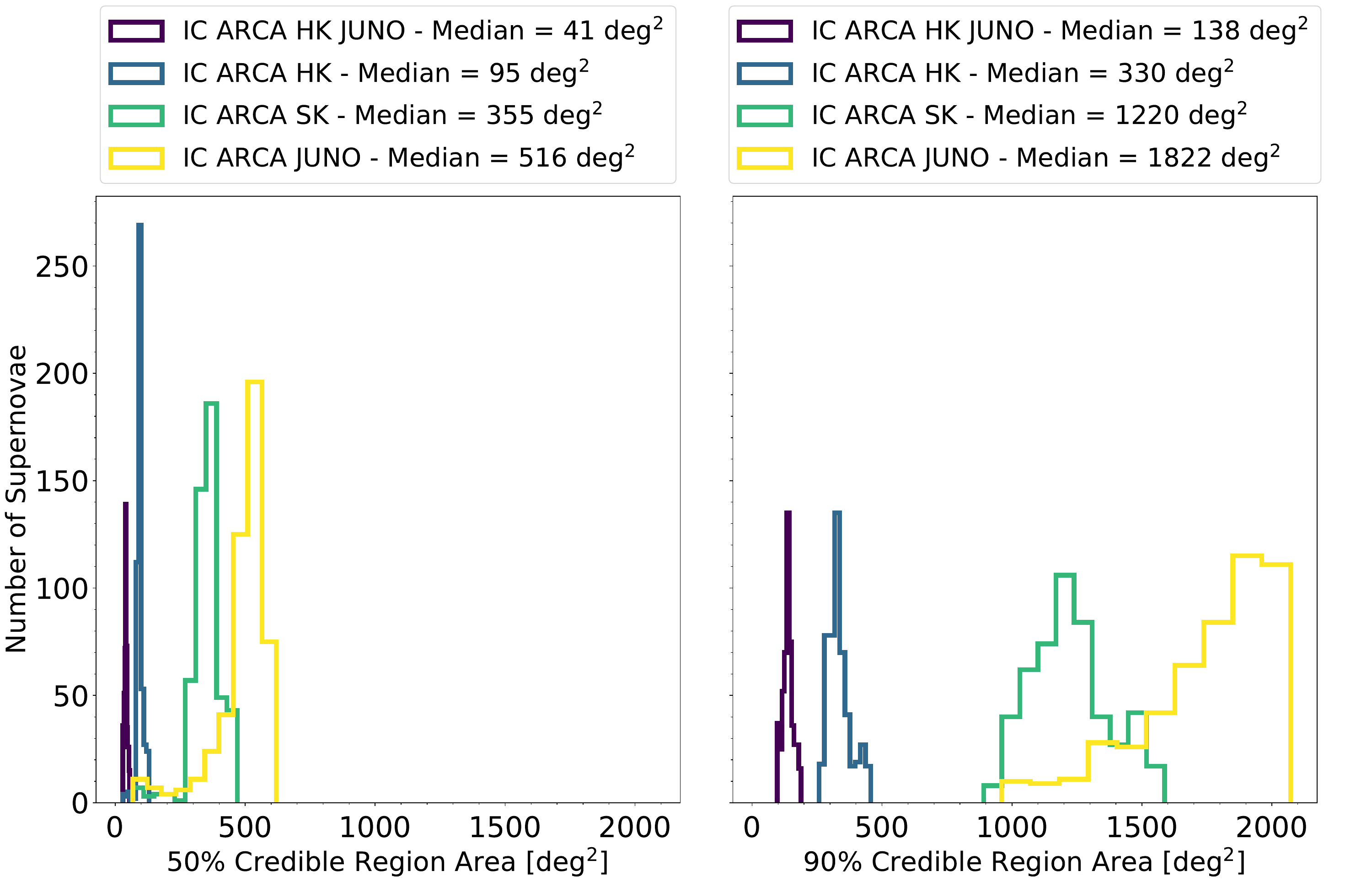}
        \caption{Distribution of the areas of the credible region for the neutrino localisation produced with the \lcm algorithm. For both panels, the dark blue curve represents the network IC-ARCA-HK-JUNO, the blue curve represents the IC-ARCA-HK network, the green curve represents the IC-ARCA-SK network, and the yellow curve corresponds to the IC-ARCA-JUNO network. The median areas are presented in the legend. The left panel corresponds to the areas of the credible region at the 50\%  level and the right one to the areas for the 90\% credible region.}
        \label{fig:cra}
    \end{figure*}

    After performing simulations of neutrino burst localizations, we obtain the corresponding skymaps that characterize the reconstructed probability regions for the source's position. For a fixed configuration of the neutrino detector network, these skymaps were found to be overall consistent in structure, exhibiting similar morphologies and angular extents across different realizations. This suggests that the dominant features of localization are primarily determined by the geometry and sensitivity of the detector array, rather than by statistical fluctuations in the simulated events. As a result, the network configuration emerges as the key factor controlling the localization performance, with only minor variations observed between individual simulations. This is visible in Fig.~\ref{fig:cra} where the distributions of the 50\% and 90\% for the networks that include the HK detector are almost vertical. An example of the contours provided by \lcm for each network configuration is shown in Fig.~\ref{fig:skm-ex} for one of the CCSN of the population. The IC-ARCA-HK-JUNO configuration provides skymaps with only a single-region skymap, where the reconstructed probability distribution is concentrated in a contiguous area. The number of detectors explains this for this network configuration, which allows breaking the spatial degeneracy that remains when only three instruments detect the neutrino burst. In the other configurations, the spatial distribution becomes multi-modal, which makes the follow-up more difficult. The two configurations that include HK provide better localizations for this detector, which is more sensitive in the energy range of the neutrinos emitted by a CCSN (around $\sim$40 MeV). The same argument holds for the IC-ARCA-SK configuration, which provides better localizations than IC-ARCA-JUNO, as SK is more sensitive in the $\sim$40MeV domain. The 50\% (left panel) and 90\% (right panel) credible region area distributions for the whole population are shown in Fig.~\ref{fig:cra}. The results are consistent with the contours presented in Fig.~\ref{fig:skm-ex} and with the examples and areas presented in \cite{coleiro2020combining}.  
    
    \subsection{Baseline results of the follow-up}
    \label{sec:baseline}
    
        \begin{figure*}
        \includegraphics[width=\textwidth]{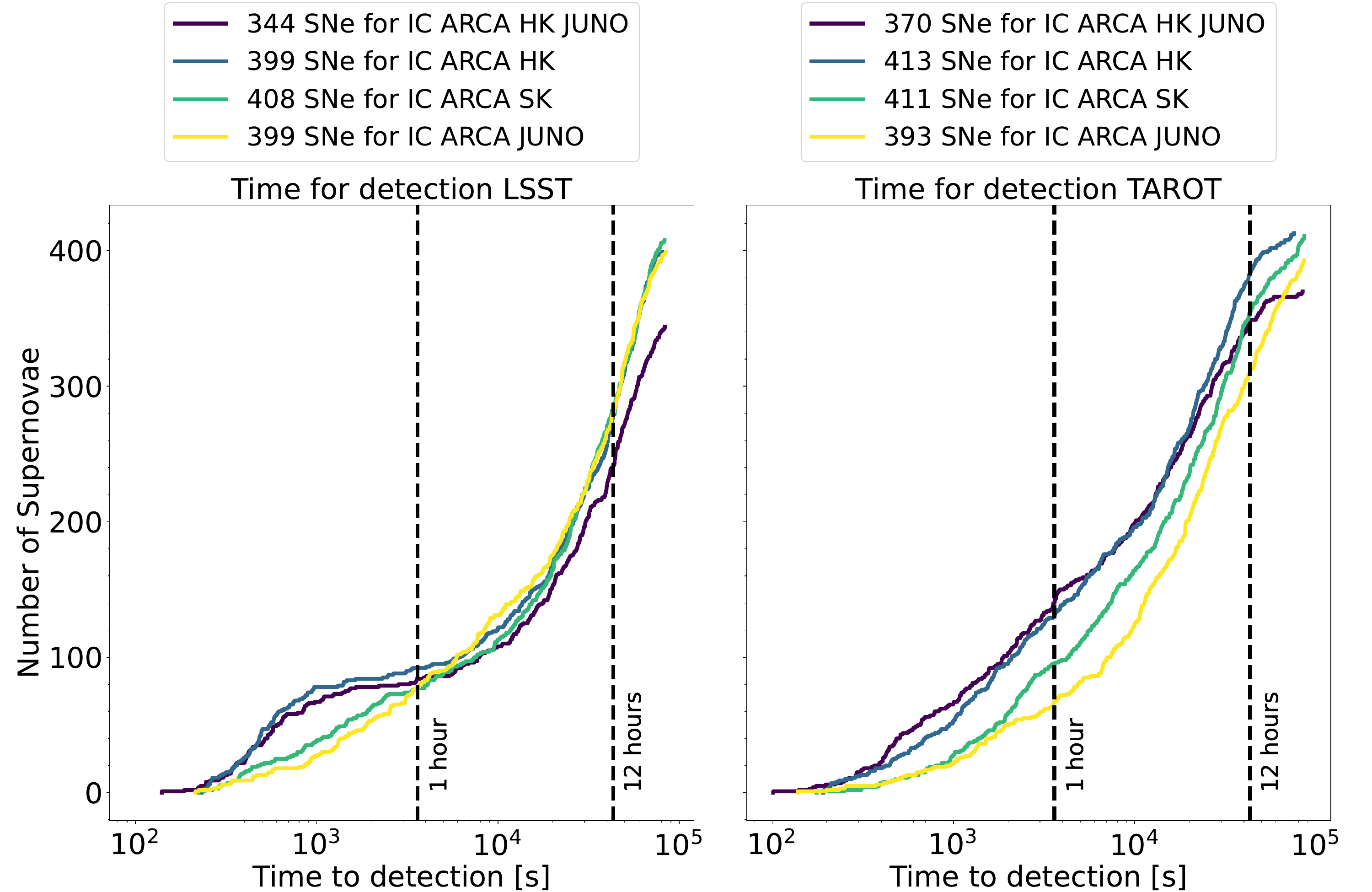}
        \caption{Cumulative distributions of the time to image the CCSN's position for LSST-VRO on the left panel and the TAROT network on the right one. Each curve corresponds to a specific neutrino network configuration: the dark blue curve represents the network IC-ARCA-HK-JUNO, the blue curve represents the IC-ARCA-HK network, the green curve represents the IC-ARCA-SK network, and the yellow curve corresponds to the IC-ARCA-JUNO network.
        }
        \label{fig:times}
    \end{figure*} 
    
    \begin{figure*}
        \includegraphics[width=\textwidth]{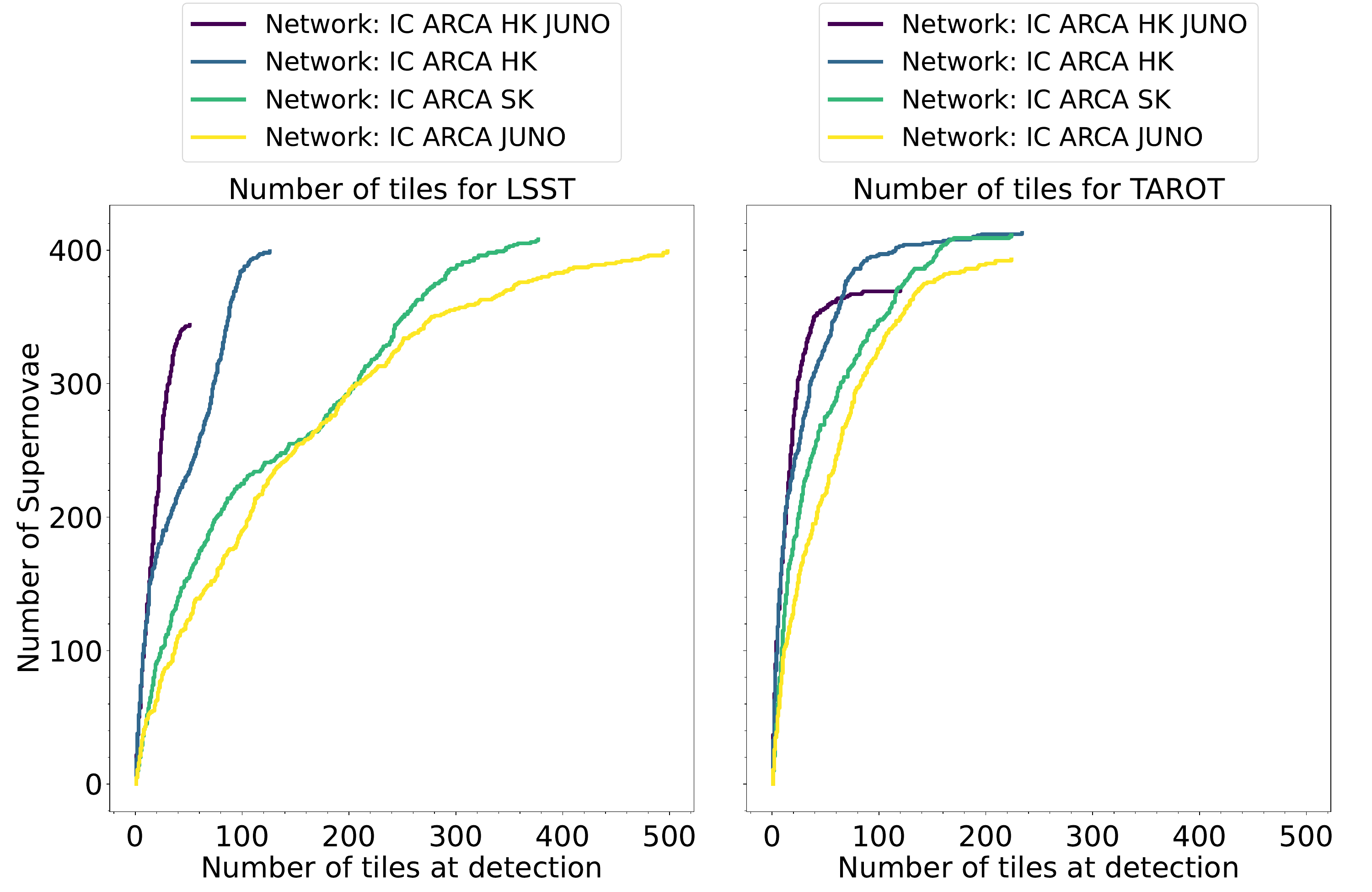}
        \caption{Cumulative distributions of the number of tiles necessary to image the CCSN's position for LSST-VRO on the left panel and the TAROT network on the right one. Each curve corresponds to a specific neutrino network configuration: the dark blue curve represents the network IC-ARCA-HK-JUNO, the blue curve represents the IC-ARCA-HK network, the green curve represents the IC-ARCA-SK network, and the yellow curve corresponds to the IC-ARCA-JUNO network.}
        \label{fig:tiles}
    \end{figure*}
    
    After producing the neutrino skymaps for each neutrino network configuration and the associated observing plan, we post-process the data to estimate the number of CCSNe observed in both the neutrino and EM regimes. \\ \noindent
    The \gwo observing plans presented in Sec.~\ref{sec:plan} are processed to locate the tile in which the injected CCSN is positioned, ensuring that the angular separation between the CCSN coordinates and the center of the identified tile is smaller than the telescope's FoV. Once the appropriate tile is confirmed, we extract the corresponding detection time of the CCSN as the observation time proposed by the scheduling of \gwo. The time to get on the tile containing the CCSN, after the randomization process described in Sec.~\ref{sec:rando}, is summarized in Fig.~\ref{fig:times}. We also extract the number of tiles to image before reaching the target's position, allowing us to evaluate the amount of observing time to dedicate to the follow-up campaign for a given facility. These results are summarized in Fig.~\ref{fig:tiles}. Regarding the TAROT network, since there are three instruments, we use the time and number of tiles of the first of the three facilities that reach the target for the estimations in Figs.~\ref {fig:times} and \ref{fig:tiles}.

    Based on Fig.\ref{fig:times} results, for all the cases where the tile containing the CCSN is imaged after the neutrino burst, the delay is always less than 24h for both LSST-VRO and TAROT. The median time is at the $\sim$7h level for LSST-VRO regardless of the neutrino network configuration, as presented in Tab.\ref{tab:med-time-baseline}. For TAROT, the median time to reach the CCSN's position is smaller for all neutrino network configurations, albeit similar for the IC-ARCA-JUNO case. This result is discussed more thoroughly in Sec. ~\ref{sec:dis}. However, there is a variation of the median time that follows the evolution of the size of the localisation contours of the neutrino network presented in Sec.~\ref{sec:nu-loc}. 

    \begin{table}[h!]
    \begin{center}
        \caption{
        Median time in hours to reach the CCSN's tile for LSST-VRO and TAROT follow-up facilities and for the four neutrino networks configurations.
       } \label{tab:med-time-baseline}
        \begin{tabular}{c||c|c}
        \hline
        \hline
        Neutrino network  & LSST-VRO & TAROT \\
        \hline
        IC ARCA HK JUNO  & 6.8h & 3.5h \\
        IC ARCA HK       & 7.2h & 4.2h \\
        IC ARCA SK       & 7.5h & 5.6h \\
        IC ARCA JUNO     & 6.8h & 6.0h \\
        \hline 
    \end{tabular}
    \end{center}
    \end{table}
    
    The median numbers of tiles to be observed before getting onto the CCSN's one for each optical facilities and the various neutrino configurations are provided in Tab.~\ref{tab:med-tile-baseline}. We observe that the TAROT facilities need to image fewer tiles than LSST-VRO for all neutrino network configurations. In particular, LSST-VRO needs to observe more than twice as many tiles as TAROT to find the CCSN for the IC-ARCA-SK and IC-ARCA-JUNO neutrino networks, which provide the largest areas. This can be explained by the slightly larger FoV of TRE, that is the TAROT instrument that is detecting the CCSN first and the use of three instruments in the TAROT network.
    
    \begin{table}[h!]
    \begin{center}
        \caption{
        Median number of tiles to observe before getting onto the CCSN's tile for LSST-VRO and TAROT follow-up facilities, and for the four neutrino networks configurations.
       } \label{tab:med-tile-baseline}
    \begin{tabular}{c||c|c}
    \hline
    \hline
    Neutrino network  & LSST-VRO & TAROT \\
    \hline
    IC ARCA HK JUNO & 16  & 13 \\
    IC ARCA HK      & 32  & 19 \\
    IC ARCA SK      & 81  & 35 \\
    IC ARCA JUNO    & 107 & 43 \\
    \hline 
    \end{tabular}
    \end{center}
    \end{table}

    \subsection{Moon's influence}
    \label{sec:moon}
    Looking at Fig.\ref{fig:times}, it is noticeable that the number of CCSNs whose position is observed by following the \gwo's observing plans is significantly lower for the configuration IC-ARCA-HK-JUNO. As this network provides the best localizations, as shown in Fig.~\ref{fig:cra}, one could expect the number of detections to be comparable or higher than for the larger skymaps, as it is easier to cover the whole credible region. This counterintuitive result is explained by the fact that when the neutrino burst is better localized, the reconstructed skymaps tend to fall in regions of the sky that overlap with the Moon's trajectory, as most of the progenitors are concentrated in the galactic center. Because these regions are more likely to be obstructed or unobservable during optical follow-up, when the Moon is bright or nearby, for example, this reduces the chance of imaging the CCSN's position. Therefore, the improved localization configuration paradoxically results in fewer detected SNe. In addition, part of this effect arises from the randomization of CCSN positions introduced in Sec.~\ref{sec:rando}. Although intended to avoid systematic biases in the follow-up campaign, this randomization can occasionally shift the simulated CCSN away from the Galactic plane and, correspondingly, away from the regions where the Moon’s trajectory affects visibility. This is affecting the large skymaps produced by the three neutrino detector configuration, as presented in Fig.~\ref{fig:skm-ex} with the blue, orange, and red contours. As they present multi-modal shapes, the randomized position is put in the region of the skymap that is not in the Galactic Plane. This can artificially prevent the overlap between the reconstructed skymap and the regions affected by the Moon's trajectory. Consequently, the improved localization combined with this positional randomization paradoxically results in fewer detected SNe. Consequently, we expect that the Moon's presence will also affect a neutrino burst detected by three detectors, and taken into account for a follow-up campaign.

\begin{figure}[h!]
        \includegraphics[width=\columnwidth]{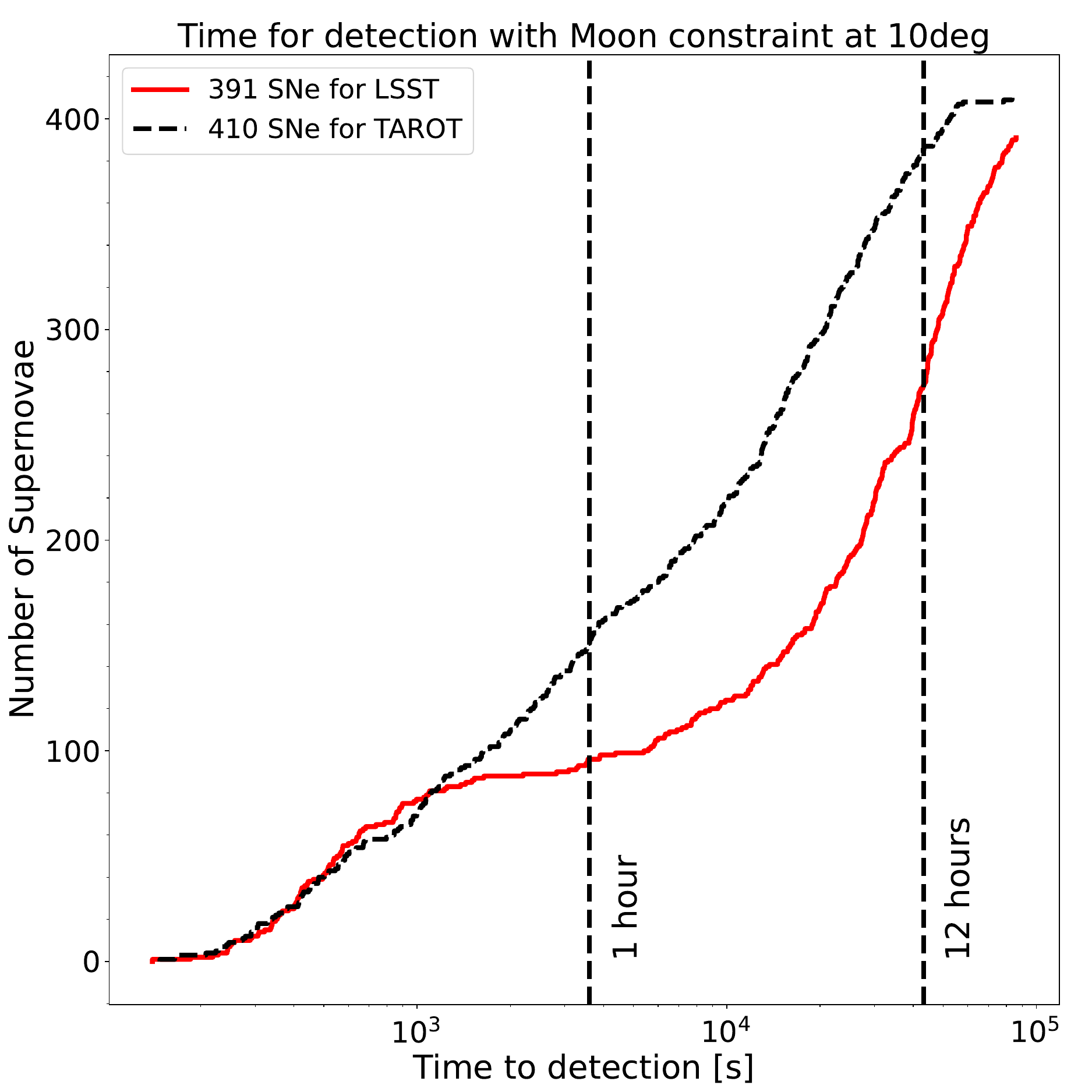}
        \caption{Cumulative distributions of the time to image the CCSN's position for the skymaps produced by the IC-ARCA-HK-JUNO network with the angular separation between the tile's center and the Moon set to 10 deg (against the 30 deg constraint used for the baseline analysis presented in Sec.~\ref{sec:baseline}). The red curve corresponds to the LSST-VRO observations, and the dashed black curve corresponds to the TAROT network results.
        }
        
        \label{fig:times-moon}
    \end{figure}

    The minimal angular separation between the Moon and a tile in a plan produced by \gwo is controlled by a parameter in the configuration file. By default, a tile has to be at least 30 deg away from the Moon's position when the tile is observed, and this is the value used to produce the results presented in Sec.~\ref{sec:baseline}. Consequently, we change the Moon constraint of \gwo to 10 degrees, which allows us to cover more of the credible region without risking direct observation of the Moon, which could damage the instrument. Then we produce the observing plan for both LSST-VRO and TAROT for the skymaps of the four-detector configuration. We process the plans in the same way as in the previous section, and we found that the CCSN's position is observed in 391 cases for LSST-VRO and 410 cases for TAROT, as presented in Fig.\ref{fig:times-moon}. These numbers are similar to the other network configurations. 

    \subsection{SBO and CCSN detections}
    \label{sec:sbo}
    In Sec.~\ref{sec:baseline}, our analysis considered only the performance of the follow-up strategy in reaching the CCSN position. However, we did not evaluate whether the CCSN was actually detected once that position was reached. In particular, the neutrino source’s location may be imaged too early, before the occurrence of the SBO, resulting in a non-detection. As the RSG progenitors have an SBO propagation time of the order of a day, as presented in Tab~\ref {tab:sbo}, none of the follow-up campaigns with an RSG progenitor is expected to lead to a detection during the first night following the neutrino detection, even if the source's position is imaged. On the contrary, we expect the detection to happen during the first night for the WR progenitors as the propagation time is of the order of a minute. The BSG progenitors can be detected rapidly after the neutrino burst, as the SBO propagation time is about an hour. The SBO can also be observed as there is a few-minute window when it is observable. However, as visible in Fig.~\ref{fig:times}, the follow-up is rapid enough to be on target before the SBO happens.
    
\begin{figure*}
        \includegraphics[width=\textwidth]{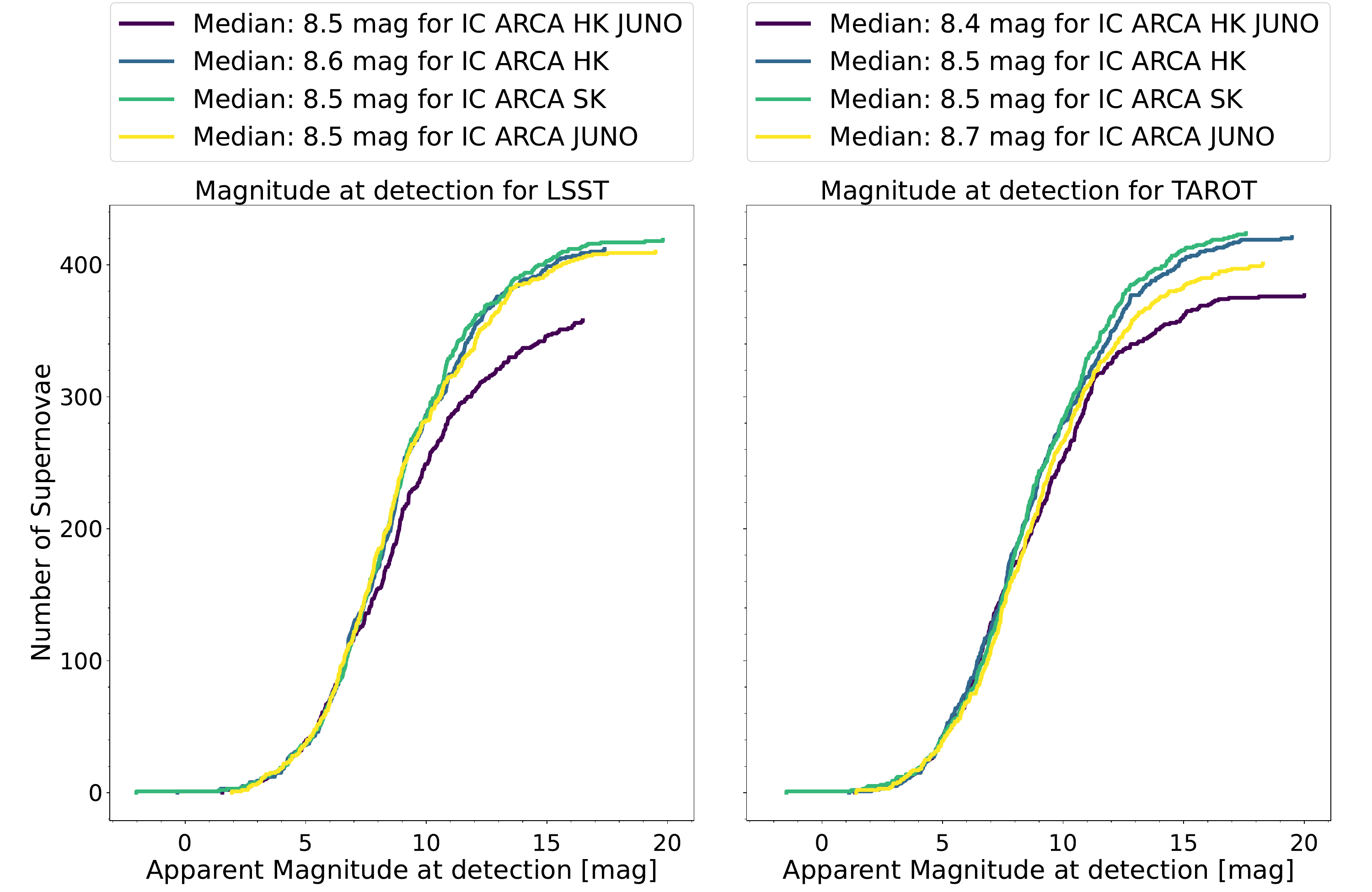}
        \caption{Cumulative distributions of the apparent magnitudes of the CCSN at the moment it is detected by LSST-VRO on the left pannel and the TAROT network on the right one. Each curve corresponds to a neutrino network configuration: the dark blue curve is for the network IC-ARCA-HK-JUNO, the blue curve is for the IC-ARCA-HK network, the green curve is for the IC-ARCA-SK network and the yellow curve corresponds to IC-ARCA-JUNO.}
        \label{fig:mag-det}
    \end{figure*}

    To evaluate the number of successful SBO detections and post-SBO detections, we compute the delay between the neutrino burst detection and the observation of the tile that contains the CCSN. We consider that the SBO is detected if the time at which the position is imaged is in the window [SBO delay, SBO delay + SBO duration], where the delay and duration are taken from Tab.~\ref{tab:sbo} and depend on the progenitor's type. For the TAROT network, we consider only the first facility to get on target. If the time to observe the CCSN's position is below the SBO delay, the optical emission is not visible as it has not happened yet, and thus, there is no detection. If the time delay exceeds the SBO delay plus SBO duration, then only the SN emission is detected, but the SBO is missed. The SN can also be missed because the observing plan produced by \gwo does not cover the CCSN, either because it is on the edges of the skymaps or because it is too close to the Moon, as detailed in Sec.~\ref{sec:moon}. Consequently, we produce observing plans for all non-detection cases for the two nights following the neutrino detection, ensuring that all SBOs occur and all CCSNs are visible. The results are summarized in Tab.~\ref{tab:det-sum}, where the counts of the SBO and post-SBO detections are shown for each follow-up night. The numbers of detections in Tab.~\ref{tab:det-sum} are consistent with the numbers visible in Fig.~\ref{fig:times}. The TAROT facilities and LSST-VRO have similar overall detection rates of the optical counterpart. However, the former instruments detect the SBO more often, except for the IC-ARCA-JUNO neutrino network. As presented in Fig~\ref{fig:cra}, this network produces the largest skymaps and thus decreases the chances of early detection of the SBO. Fig.~\ref{fig:mag-det} shows the magnitude at which the neutrino source is detected for the various neutrino networks. The median magnitude at which the SN is observed is $\sim$8.5 mag for all the networks and for both LSST-VRO and TAROT. However, for the latter, a marginal number of detections happen below the limiting magnitudes presented in Tab.~\ref{tab:gwemopt_param}. In particular, for the TRE facility, which has a depth of 16 mag with the exposure time used in Sec.~\ref{sec:plan}, this might lead to weak detections in the frame. These high magnitudes are explained by two cases: the galactic absorption and the early detections that are allowed by the TAROT network, which happen at the very early stages of the EM emission when it is still faint, as it is visible in Fig.~\ref{fig:lc-ex}. 
    
    \begin{table*}[h!]
    \begin{center}
        \caption{
        Summary of the detections for the different neutrino networks and for the two optical follow-up facilities. For each cell, the first line provides the number of SBO detected, the second line the number of SN detection without the SBO and the third line is the total number of detections over the whole follow-up. In the first two lines of a row, each number corresponds to the night of the follow-up when the progenitor is detected by LSST-VRO or a TAROT instrument.
       } \label{tab:det-sum}
    \begin{tabular}{c||c|c}
    \hline
    \hline
    Neutrino network  & LSST-VRO & TAROT \\
    \hline
    IC ARCA HK JUNO & SBO = 0, 8, 0  & SBO = 2, 14, 0 \\
                     & CCSN = 90, 163, 97 & CCSN = 88, 112, 161 \\
    Total number of detections & 358 & 377\\
    \hline
    IC ARCA HK      & SBO = 0, 9, 0  & SBO = 2, 16, 0 \\
                    & CCSN = 104, 194, 105 & CCSN = 102, 142, 159 \\
    Total number of detections & 412 & 421 \\
    \hline
    IC ARCA SK      & SBO = 0, 11, 0  & SBO = 2, 17, 0 \\
                    & CCSN = 109, 206, 93 & CCSN = 112, 155, 138 \\
    Total number of detections & 419 & 424\\
    \hline
    IC ARCA JUNO    & SBO = 2, 8, 0  & SBO = 0, 10, 0 \\
                    & CCSN = 109, 180, 111 & CCSN = 108, 161, 122 \\
    Total number of detections & 410 & 401 \\
    \hline 
    
    \end{tabular}
    \end{center}
    \end{table*}

\section{Discussions}
    \label{sec:dis}
    The simulations we described in the previous sections provide a quantitative evaluation of the expected performance of optical facilities following a neutrino-detected CCSN. Based on these results, we now examine the observational limitations, instrumental characteristics, and coordination requirements that would ensure an operational follow-up strategy.
    
    \subsection{Overview of the neutrino follow-up strategy}
    The blind search phase of the follow-up campaign is expected to extend over several nights, as the emergence of the optical counterpart depends strongly on the progenitor type and explosion properties. As a result, optical facilities will need to dedicate a substantial fraction of observing time across multiple nights to ensure adequate temporal coverage of the potential event. The main target for early observation is to detect the SBO phase, however, the fraction of events with a detected SBO in our simulated optical follow-up campaigns remains small. The emission is found in less than 5\% of the cases as visible in Tab.~\ref{tab:det-sum}. Moreover, as most of the SBO emission occurs in the soft X/UV domain, a follow-up campaign to detect the neutrino burst would benefit from coordinating with space-based facilities such as Neil Gehrels Swift Observatory \citep{2004ApJ...611.1005G} and SVOM \citep{2016arXiv161006892W}, which can perform tilling observations in the UV bands. However, the production of the observing plans has to take into account the slew constraints for making the pointings. In particular, the multimodal skymaps produced by three-detector configurations, as shown in Fig.~\ref{fig:skm-ex}, are challenging to tile, as they require significant observatory motion in addition to the timing constraint for early detection of a short-lived transient, such as the SBO. 
    An additional operational constraint that we identified and must consider in real-time follow-up campaigns is the likely proximity of the target field to the Moon, particularly in cases where a burst is detected with IC, ARCA, HK, and JUNO. For typical transient searches, observations within 30 degrees of the Moon are avoided to minimize background and prevent instrumental damage; however, for neutrino-triggered CCSN follow-ups, relaxing this constraint may be necessary to increase the chance of early detection of the EM counterpart as presented in Sec.~\ref{sec:moon}. Observations initiated at angular separations as close as 10 degrees could be acceptable, provided that instrument safety is maintained. This requirement might prove challenging for large facilities, such as LSST–VRO, which require strict illumination constraints on their optics, but smaller telescopes, like TAROT, are generally more flexible. Given the expected optical brightness of nearby CCSNe, especially during the first few days post-explosion, the increased background at such angular distances is unlikely to significantly hinder detection. Incorporating adaptable Moon avoidance policies into follow-up scheduling would thus improve sky coverage and detection probability without compromising data quality.

    The results presented in this study are conservative, as they rely on the randomization procedure described in Sec.~\ref{sec:rando}. This randomization can place simulated CCSNe outside the Galactic plane, where CCSNe are expected to occur (Sec.~\ref{sec:dis}), thereby increasing the number of required telescope tiles and the time needed to reach the target, and reducing the detection efficiency. Consequently, this approach tends to underestimate the performance of an actual follow-up campaign focused on Galactic-plane regions. Nevertheless, the randomization is necessary to avoid underestimation bias: without it, reconstructed CCSN positions would systematically lie in the most probable regions of the \lcm skymaps, leading to overly optimistic follow-up times and sky coverage. By allowing source positions to vary across the skymap, this method provides a more robust and unbiased—though conservative—assessment of the observing strategy.
    
    \subsection{Role of small robotic instrument}
    The TAROT network, along with other small-aperture but wide FoV robotic telescopes, is ideally suited for the initial blind search phase following a neutrino trigger. These instruments are designed for rapid response and wide-field coverage, characteristics that are essential for the efficient exploration of large localization regions. Initially designed for the follow-up of gamma-ray burst (GRB) afterglows, TAROT telescopes can rapidly tile the sky with minimal human intervention, providing early-time optical imaging within minutes of an alert \citep{2024A&A...682A.141T}. This capability makes them highly valuable in the context of CCSN neutrino alerts, where localization uncertainties remain significant and time is critical. Moreover, the use of multiple, geographically distributed small facilities increases temporal coverage and mitigates weather-related risks. Therefore, TAROT and similar networks could play a crucial role in the initial detection of optical counterparts in a multi-messenger campaign related to a CCSN neutrino burst.

    \subsection{Current LSST-VRO follow-up strategy}
    The LSST–VRO collaboration has proposed an observational strategy for responding to a Galactic CCSN neutrino trigger in \cite{2024arXiv241104793A}. It assumes a localization region of approximately 25 square degrees for a detection with the SK detector, which is achieved for a version of SK that includes gadolinium, as presented in Fig.9 of \cite{10.1093/mnras/stw1453}. If SK keeps its current configuration, the resolution is more at the $\sim$100 deg$^2$. These estimations are based on the individual pointing capability of the instrument and not on signal triangulation as we presented in Sec.~\ref{sec:nu-loc}. The plan considers a 100 square degrees tilling for an alert if no counterpart is discovered, and this will be repeated for 1-2 days post neutrino trigger. This approach is consistent with the one we used in Sec.~\ref{sec:sbo} and is expected to cover the propagation time of the SBO of RSG progenitors. However, if the localisation provided by the SNEWS collaboration relies on triangulation rather than the individual capability of a detector, the 100 square degrees limit may be too restrictive. As visible in Fig.~\ref{fig:cra}, the typical 90\% credible region is more at the 1,000 square degrees for the network configuration that includes SK. This limit is more suited for a detection with the HK instrument, which is expected to operate in the same period as LSST-VRO.

\section{Conclusion}

    In this work, we explored strategies that can be implemented to enable multi-messenger detections of a CCSN in both the neutrino and optical regimes. Such a joint detection is currently expected for an event that would occur in the MW or one of its close satellites, due to the sensitivities of the neutrino detectors. Considering the galactic rates of CCSNe, which are approximately one per century, the detection of a SN with neutrinos is considered an exceptional event worthy of dedicating a significant amount of observing time. However, these resources have to be coordinated to maximize the chance of early EM detection. To investigate the planning of the follow-up, we modelled a population of galactic CCSN with a spatial distribution following a double exponential distribution. We considered three different progenitors (WR, WSG, and RSG) with varying fractions in the MW, as estimated from optical survey results. For the optical emission, we used two template lightcurve models corresponding to Type IIp (RSG and BSG progenitors) and Ib/c (WR progenitors). We also included an SBO emission for the RSG and BSG that are observable in the optical bands, even if this domain is not the preferred one for this EM counterpart. It is not simulated for the WR as these emissions happen too rapidly to be detectable in a planned search campaign.

    As neutrinos are the first messengers to be emitted during the explosion, they are expected to initiate the observation. Consequently, we considered different neutrino detectors operating as a network, as is done within the SNEWS collaboration. The networks we simulated include already existing instruments such as SK, KM3NeT/ARCA and IC, along with upcoming ones, such as JUNO and HK. The neutrino burst is assumed to be detected by the whole network that can then triangulate the source and produce a skymap that is publicly shared and used by the optical facilities to search for the EM counterparts. The triangulation is performed using the \lcm method, which employs a chi-square estimator and the cross-correlation of neutrino lightcurves to compute the time delay between all pairs of detectors in the network and estimate the probability distribution on the sky. The skymaps produced with the \lcm algorithm are used as an input for \gwo to produce observing plans. This is the software currently used routinely for the follow-up of GW events to plan the observations of various optical facilities. We used LSST-VRO and the TAROT network as the instruments participating in the campaign to find the neutrino source. The large FoV and depth of LSST-VRO offer a sensitivity to faint or highly extinguished events. At the same time, the robotic and distributed nature of the TAROT telescopes enables rapid coverage of the neutrino skymaps. The simulations we described show that both facilities are capable of detecting the optical counterparts of neutrino-detected CCSNe with similar success rates. In addition, TAROT achieves comparable detection efficiency while requiring fewer tiles than LSST-VRO, demonstrating that modest, fast-response telescopes can play a significant role in the prompt identification of nearby CCSNe. Moreover, our results emphasize the value of rapid communication and automated scheduling systems in minimizing latency between neutrino detection and optical response, as well as the complementary roles that large-aperture and small, robotic telescopes can play in neutrino-triggered CCSN searches. 
    
    The overall success of neutrino-triggered CCSN follow-up efforts — defined here as the early optical detection of the supernova, and potentially even the detection of the SBO — will strongly depend on the efficiency of observational planning and coordination among participating facilities. Rapid communication between neutrino observatories and the optical community, along with automated scheduling tools such as \gwo, is crucial for optimizing telescope time and ensuring timely observations of the skymaps. Coordinated coverage between large and small facilities can substantially improve the detection chances, particularly when the localisation regions span tens to thousands of square degrees.
    


\section*{Competing Interests}
The authors have no relevant financial or financial interests to disclose.



\section*{Data Availability}
The data that support the results of this study are available from the corresponding author, P.-A. Duverne upon request, and will also be part of a data release in 2026.

\section*{Code Availability}
The code is available on github: \url{https://github.com/PAduverne/nextsngal}.

\begin{acknowledgements}
This work was supported by the Programme National des Hautes Énergies (PNHE) of CNRS/INSU co-funded by CNRS/IN2P3, CNRS/INP, CEA and CNES. 
PAD thank Pr Green for his help using the \texttt{BAYESTAR} dustmaps, Dr. S. Anand for the insight on the relevant configuration to use \gwo for LSST-VRO and Dr. S. El Hedri for her insights on the time delay between the neutrino burst and the EM emission.
MWC acknowledges support from the National Science Foundation with grant numbers PHY-2117997, PHY-2308862 and PHY-2409481.
\end{acknowledgements}


\bibliographystyle{aa}
\bibliography{references}

\end{document}